\documentclass[prl,aps,amsmath,twocolumn,showpacs,floatfix]{revtex4}
\usepackage{graphicx}
\begin{document}
%\draft
\title{Percolation with excluded small clusters and Coulomb blockade in a granular system}
\author{A.~S.~Ioselevich  and D.~S.~Lyubshin}
\affiliation{Landau Institute for Theoretical Physics RAS,  117940  Moscow, Russia,\\
Moscow Institute of Physics and Technology, Moscow 141700,
Russia.}
\date{\today}

\date{\today}

\begin{abstract}
We consider dc-conductivity $\sigma$ of a mixture of small
conducting and insulating grains slightly below the percolation
threshold, where finite clusters of conducting grains are
characterized by a wide spectrum of sizes. The charge transport is
controlled by tunneling of carriers between neighboring conducting
clusters via short ``links'' consisting of one insulating grain.
Upon lowering temperature small clusters (up to some $T$-dependent
size) become Coulomb blockaded, and are avoided, if possible, by
relevant hopping paths. We introduce a relevant percolational
problem of next-nearest-neighbors (NNN) conductivity with excluded
small clusters and demonstrate (both numerically and analytically)
that $\sigma$ decreases as power law of the size of excluded
clusters. As a physical consequence, the conductivity is a
power-law function of temperature in a wide intermediate
temperature range. We express the corresponding index through
known critical indices of the percolation theory and confirm this
relation numerically.

\end{abstract}

\maketitle

A mixture of metallic and insulating grains is an important
practical system appearing in numerous applications, such as
ceramics, catalysts, and powders. It is also a useful model,
describing many realistic composite materials (see \cite{book}).
Normally the properties of this system are described on the
classical level: by the effective medium theory \cite{kirkpatrick}
(away from the percolation threshold), or by the percolation
theory \cite{stauffer-aharony,bunde-havlin} (near the threshold).
For traditional ``old-fashioned'' mixtures (with the sizes of
grains $a$ on the scale of microns), the classical approach is
sufficient. However, in  recent years a special attention was
drawn to nanocomposite materials with characteristic $a\sim 10\;
nm$ \cite{nanocomposites}. For such small grains essentially
quantum effects should become important, especially on the
insulating side of the percolation transition: tunneling through
insulating grains and the Coulomb blockade effect (see, e.g.,
\cite{Devoret}).

\begin{figure}
\includegraphics[width=0.9\columnwidth]{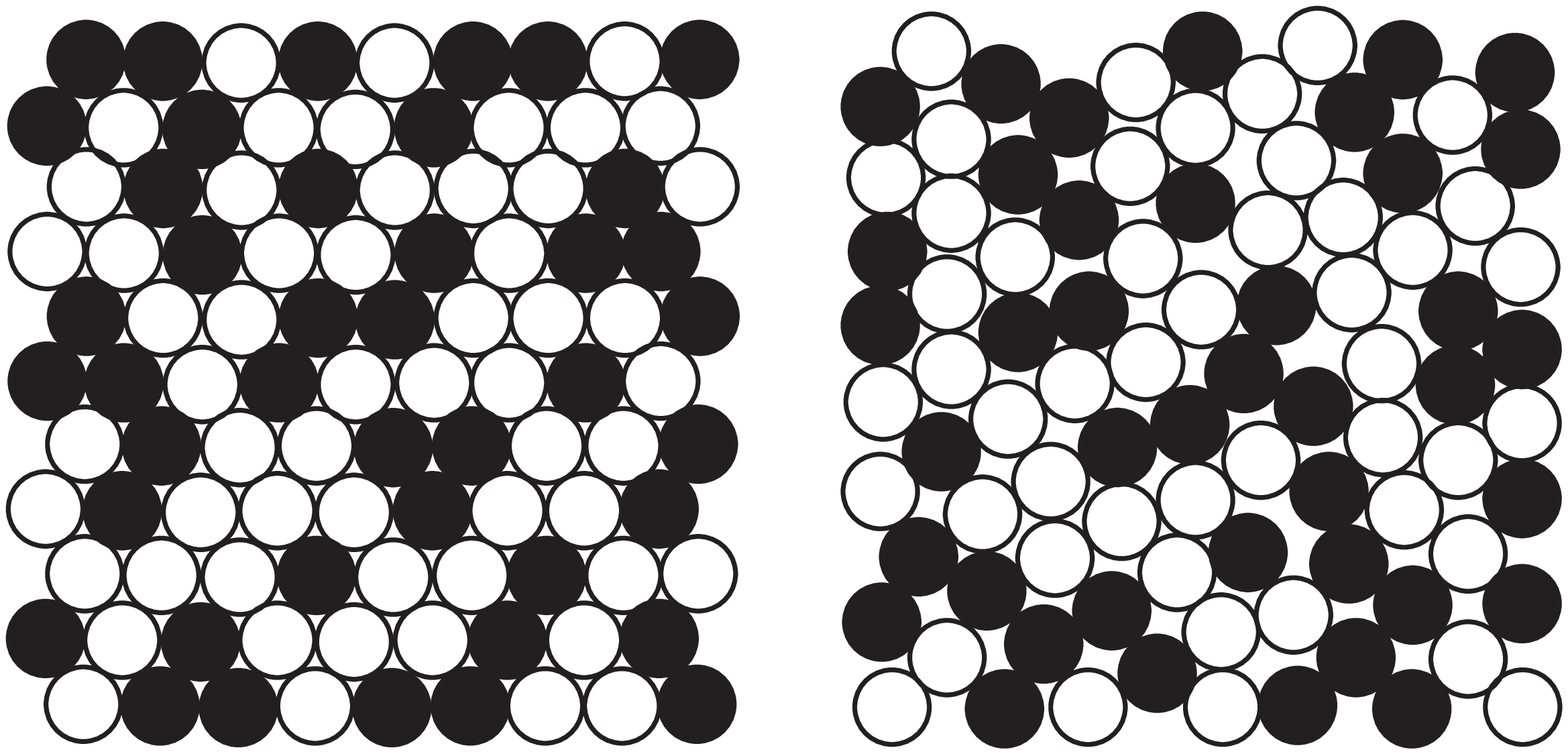}
\caption{A  mixture of conducting grains (shown black) and
insulating ones (white). Left panel: regular dense packing; right
panel: topologically disordered random packing.
 }
 \label{mixture}
\end{figure}

In this paper, for definiteness, we consider a  mixture of
spherical grains of the same diameter $a$, the fraction $x$ of
them being metallic, while the fraction $1-x$ being insulating
(Fig.\ref{mixture}). Every pair of adjacent conducting grains is
supposed to establish a good electric contact with dimensionless
conductance $G\gg 1$, but still the resistance $1/G$ of such a
contact is larger than the resistance of a conducting grain
itself, so that the latter will be altogether neglected in what
follows.

If one ignores the transport through insulating grains, then the
macroscopic dc-conductivity $\sigma$ of the system identically
vanishes below the percolation threshold (for $x<x_c$), so that
the system undergoes a sharp metal-insulator transition (see
\cite{stauffer-aharony,bunde-havlin}). The infinite metallic
cluster penetrating the entire system and carrying the dc current
does not exist below the threshold: there are only finite metallic
clusters embedded into the insulating matrix. The value of $x_c$
depends on the particular arrangement of the mixture. For example,
for the regular densely packed 3d array of identical spheres (see
Fig.\ref{mixture}, left panel) $x_c\approx 0.2$
\cite{bunde-havlin}, while for the 3d random packing (right panel)
$x_c\approx 0.35$ \cite{random}. The critical exponents
corresponding to the percolation transition are, however,
universal and depend only on the dimensionality of the system $d$.

\begin{figure}
\includegraphics[width=1\columnwidth]{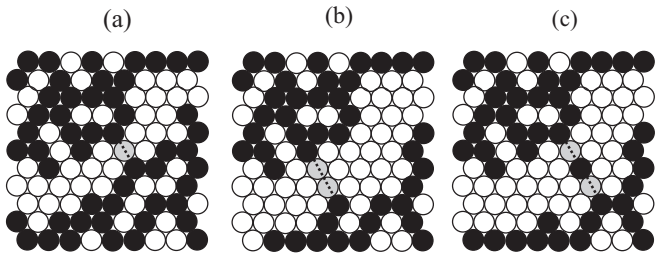}
\caption{Examples of paths (shown grey) through the insulator,
connecting two large metallic clusters. (a) A single-grain bridge
(a link). The links remain effective connectors of metallic
clusters even at lowest temperatures; (b) A two-grain bridge. In
the tunneling mode these and longer bridges do not contribute to
the conductivity even at $T>E_C$; (c) A two-link chain path with
an intermediate small (one-grain) metallic cluster. The paths
involving small clusters are the first to be frozen out at
$T<E_C$.
 }
 \label{3links}
\end{figure}

 Thus, to obtain a finite
conductivity for $x<x_c$ one should necessarily take into account
processes in which  electrons traverse insulating grains. Let  $g$
be the dimensionless conductance of the shortest possible
insulating bridge (consisting of a single insulating grain
separating two conducting clusters, see Fig.\ref{3links}a). In
what follows, we call these shortest  bridges ``links''. Though
usually $g$ is exponentially small, its finiteness smears the
metal-insulator transition, and the macroscopic conductivity of
the system below the threshold becomes nonzero.

If insulating grains are large and/or the temperature is
relatively high, then electrons penetrate insulating grains due to
activation mechanism (so that $g=g_{\rm act}\propto\exp\{-W/T\}$,
where $W$ is the energy barrier for activation of a carrier in the
insulator). For this incoherent mechanism the conductance
$g^{(n)}$ of a bridge consisting of $n$ insulating grains (see
Fig.\ref{3links}b for a $n=2$ bridge) is
\begin{eqnarray}
 g^{(n)}=g_{\rm act}/n,  \label{class-act}
\end{eqnarray}
and all such bridges with not very large $n$ give comparable
contributions to the effective conductance between two metallic
clusters. Under the classical activation conditions the system is
adequately described by a network made of randomly distributed
classical conductances of two types: $G\gg 1$ and $g\ll 1$, and
the percolation through only large conductances is established at
$x=x_c$.
 The macroscopic conductivity $\sigma$ (measured in the units of $(e^2/\hbar)a^{2-d}$) of such a random network
near the threshold is described by the formulas
\cite{es76,bunde-havlin}
\begin{eqnarray}
\sigma_{\rm ins}^{(0)}\sim
  g(x_c-x)^{-s}, & \quad   \Delta^{(0)}\ll x_c-x\ll 1,\label{class1}\\
\sigma_{\rm met}^{(0)}\sim G(x-x_c)^{\mu}, & \quad \Delta^{(0)}\ll
x-x_c\ll 1, \label{class3}
\end{eqnarray}
where  $\mu$ is the universal conductivity exponent  and $s$ is
the universal   dielectric permeability exponent (see Table
\ref{table1}). Here
\begin{eqnarray}
\Delta^{(0)}=(g/G)^{1/(\mu+s)} \label{class3kj}
\end{eqnarray}
is the width of the crossover region where the contribution of the
processes involving the insulating grains becomes comparable with
the conductivity of the infinite cluster:
\begin{eqnarray}
\sigma_{\rm ins}^{(0)}(x_c-\Delta^{(0)})\sim \sigma_{\rm
met}^{(0)}(x_c+\Delta^{(0)}).\label{class3kgg}
\end{eqnarray}
 Within the crossover region
\begin{eqnarray}
\sigma_{\rm cross}^{(0)}\sim g^{\mu/(\mu+s)}G^{s/(\mu+s)},\qquad
|x_c-x|\lesssim \Delta^{(0)}. \label{class3io}
\end{eqnarray}
 The formula \eqref{class3} does not
involve $g$, since the current manages to totally  avoid links in
this range of concentrations, so that the potential drops are
distributed over the metallic infinite cluster, while the
insulating component of the mixture does not play any role in
dc-conductivity.  Hence the classical formula \eqref{class3} is
valid for description of the mixture independent on the nature of
the processes in the insulator. That is not the case for the
formula \eqref{class1} that describes the regime where the
potential drops occur only on the insulating links between the
metallic clusters, while the latter are essentially equipotential.

\begin{table}[h]
\begin{tabular}{||c||c|c|c|c|c|c|c|c|c||}
\hline $d$ & $\nu$  & $d_f$ & $\tau$  & $\mu$ & $\tilde{\zeta}$ &
$s$ & $\theta$ & $\Theta$ & $\Theta'$\\
\hline \hline d=2 & 4/3  & 91/48 & 187/91   & 1.30 &  0.974& 1.30 &  0.123  & 0.14 & 0.07\\
\hline d=3 & 0.875  & 2.524 & 2.32  & 2.14 &  1.3 & 0.74 &  0.28 & 0.38 & 0.28\\
\hline
\end{tabular}
\caption{Numerical values of some critical
exponents}\label{table1}
\end{table}

For small grains and/or low temperatures, however, the tunneling
mechanism  dominates the transport via insulating grains, so that
the conductance of the link is $g=g_{\rm tun}\propto\exp\{-2\kappa
a\}$, where $\kappa$ is the tunneling decrement of the electronic
wave function in the insulator \cite{comment}. In the tunneling
mode the effective conductance of an insulating bridge of $n$
grains is
\begin{eqnarray}
 g^{(n)}=A_n g_{\rm tun}^n,  \label{quant-tun}
\end{eqnarray}
where $A_n$ is a model-dependent coefficient (typically $A_n\ll 1$
for $n>1$). An electron in this case has to tunnel through a
sequence of $n$ insulating grains without any intermediate stop,
and the corresponding  tunneling amplitude exponentially decays
with the length of the bridge.

Under the tunneling conditions the model of a network with two
types of conductances is generally inapplicable. Instead of one
smeared percolation transition at $x=x_c$, a {\it cascade} of
smeared transitions at $x=x_c^{(m)}$ arises, $x_c^{(m)}$ being the
threshold for percolation in the model where not only nearest
neighbors but also next nearest and farther---up to $m$-th nearest
neighbors---are assumed to be connected. In the range
$x_c^{(m-1)}<x<x_c^{(m)}$ the conductivity $\sigma\sim g_{\rm
tun}^{(m)}$ is dominated by bridges of length $m$. The crossover
between the adjacent ranges at $x=x_c^{(m)}$ is described by the
formulas similar to \eqref{class1}, \eqref{class3} with $g_{\rm
tun}^{(m)}$ standing for $g$ and $g_{\rm tun}^{(m-1)}$ standing
for $G$.  Note that this cascade of smeared phase transitions has
the same origin as the cascade-like concentration dependence of
the temperature $T_g(x)$ of the spin-glass transition  in those
dilute magnetic alloys where
 the antiferromagnetic exchange interaction of
localized magnetic ions  exponentially decreases with distance
\cite{semimagnets}.

Since we are interested in the case of very small grains and low
temperatures, in the present paper we will adopt the tunneling
scenario and concentrate on the  behavior of $\sigma$ just below
the principal threshold $x_c^{(1)}\equiv x_c$, in the
insulator-dominated regime where the direct nearest-neighbor (NN)
percolation via metal grains is impossible but the
next-nearest-neighbor (NNN) percolation (which allows paths
involving links, solitary insulating grains) still exists. Even
without taking the Coulomb effects into account, the
 conductivity $\sigma_{\rm ins}$ in the tunneling case should be somewhat
suppressed compared to the  classic percolational result
\eqref{class1}: $\sigma_{\rm ins}<\sigma_{\rm ins}^{(0)}$. Indeed,
in the tunneling regime the contributions to the conductance
between metallic clusters coming from the insulating bridges with
$n>1$ are suppressed, and only the contribution of links survives.
Numerical simulations, however, demonstrate that $\sigma_{\rm
ins}$ differs from $\sigma_{\rm ins}^{(0)}$ only in a somewhat
smaller numerical coefficient (see below), while the index $s$
remains the same:
\begin{eqnarray}
\sigma_{\rm ins}(x)=c\sigma_{\rm ins}^{(0)}\sim
  g(x_c-x)^{-s}, \qquad c<1. \label{class3ee}
\end{eqnarray}
This means that the problem with tunneling conductances remains in
the same universality class as the standard one. The insensitivity
of the index $s$ to suppression of insulating long bridges
indicates that the intercluster conductance in the case of
activation scenario is dominated by relatively short bridges of
$n\sim 1$ insulating grains anyway, while long bridges of length
$n\gg 1$ are irrelevant. Note that only the contribution of these
very long bridges could be responsible for the difference in the
values of $s$ for tunneling and activation cases (if there were
any).

Slightly below the percolation threshold the system consists of
finite metallic clusters characterized by a wide spectrum of
numbers of grains $n_i$, sizes $r_i$, conductances ${\cal G}_i$,
and charging energies $E_i$. The distribution function for
clusters with $n$ grains is well known (see, e.g.,
\cite{bunde-havlin}):
\begin{eqnarray}
N(n)\sim n^{-\tau}f(n/n_c),\qquad n_c\sim \xi^{d_f},\qquad
f(x)\sim e^{-x}, \label{clust1}
\end{eqnarray}
where $d_f$ is the fractal dimension of the infinite cluster, and
$\xi\sim a(x_c-x)^{-\nu}\gg 1$ is the correlation length. The
clusters with the size $r\sim \xi$ and $n\sim n_c$ are commonly
called {\it critical clusters}. The majority of grains belong to
small clusters with $n\sim 1$ sites. The necessary exponents are
given in the Table \ref{table1}.

In the range $1\ll n\ll n_c\sim\xi^{d_f}$ the clusters are
fractals, their physical characteristics are strongly fluctuating.
However, one can estimate  the corresponding average values for a
fixed cluster size $n$.

 An average  dimensionless conductance across the cluster
$i$ of fixed size $n$ in the range $1\ll n\ll n_c$ can be
estimated using standard expressions for conductivity of fractal
objects (see \cite{bunde-havlin}). The result is:
\begin{eqnarray}
\mathcal{G}(n)\equiv \langle \mathcal{G}_{i}\rangle_{n={\rm
const}}\sim Gn^{-\tilde{\zeta}/d_f},\quad \tilde{\zeta}\equiv
2-d+\mu/\nu.\nonumber
\end{eqnarray}
The clusters with $n_i<n_{\rm ext}$, where
\begin{eqnarray}
n_{\rm ext}\sim G^{d_f/\tilde{\zeta}}\gg 1, \label{cross-scale}
\end{eqnarray}
typically have large conductances $\mathcal{G}_i>1$ and can be
treated as structureless {\it point-like objects}. Clusters with
$n_i>n_{\rm ext}$ (and with $\mathcal{G}_i<1$) are, however,
essentially {\it extended objects}. In this paper we consider only
the case
\begin{eqnarray}
n_{\rm ext}>n_c, \label{cross-scale1}
\end{eqnarray}
when practically all existing clusters are point-like ones. The
opposite case will be considered in a separate publication
\cite{feigel}.

An average charging energy of  clusters with fixed $n$ can be
estimated from the scaling law for the average screened Coulomb
interaction
\begin{eqnarray}
\langle U(r)\rangle\sim E_C^{(0)}(a/r)^{(\nu+s)/\nu},
\label{hyp5ee}
\end{eqnarray}
where
\begin{eqnarray}
E_C^{(0)}\sim e^2/a \label{scale1}
\end{eqnarray}
is the charging energy of a smallest one-grain cluster. The
expression \eqref{hyp5ee} is valid in the intermediate range $1\ll
r\ll\xi$ and gives a correct interpolation between the unscreened
interaction $U(r)\sim E_C^{(0)}/r$ at $r\sim 1$ and the screened
one $\langle U(r)\rangle\sim E_C^{(0)}/r\tilde{\epsilon}$ at
$r>\xi$ (here $\tilde{\epsilon}\sim (x_c-x)^{-s}$ is the static
long-range dielectric constant of the system
\cite{es76,EfrosShklovsky78}). As a result,
\begin{eqnarray} E_C(n)\equiv\langle E_i\rangle_{n={\rm
const}}\sim E_C^{(0)}n^{-(\nu+s)/\nu d_f}. \label{spread5pt}
\end{eqnarray}

For temperatures $T>E_C^{(0)}$ the Coulomb blockade effects are
completely suppressed, and the  expression \eqref{class3ee}, very
similar to the  classic result \eqref{class1}, is valid.

In the range $E_C^{\rm (cr)}<T<E_C^{(0)}$ many clusters (those
with $E_i>T$) are already frozen. The freezing condition may be
rewritten as $n<\tilde{n}(T)$, where
\begin{eqnarray}
\tilde{n}(T)\sim (E_C^{(0)}/T)^{\nu d_f/(\nu+s)}.
\label{spread5ptp}
\end{eqnarray}

How does the conduction proceed in such a system? Clearly, an
electron would prefer to travel via large clusters and  to avoid
small frozen clusters. Is the NNN-percolation still possible after
the exclusion of small clusters with $n<\tilde{n}$? How does the
exclusion affect the conductivity $\sigma$?

\begin{figure}
\includegraphics[width=0.9\columnwidth]{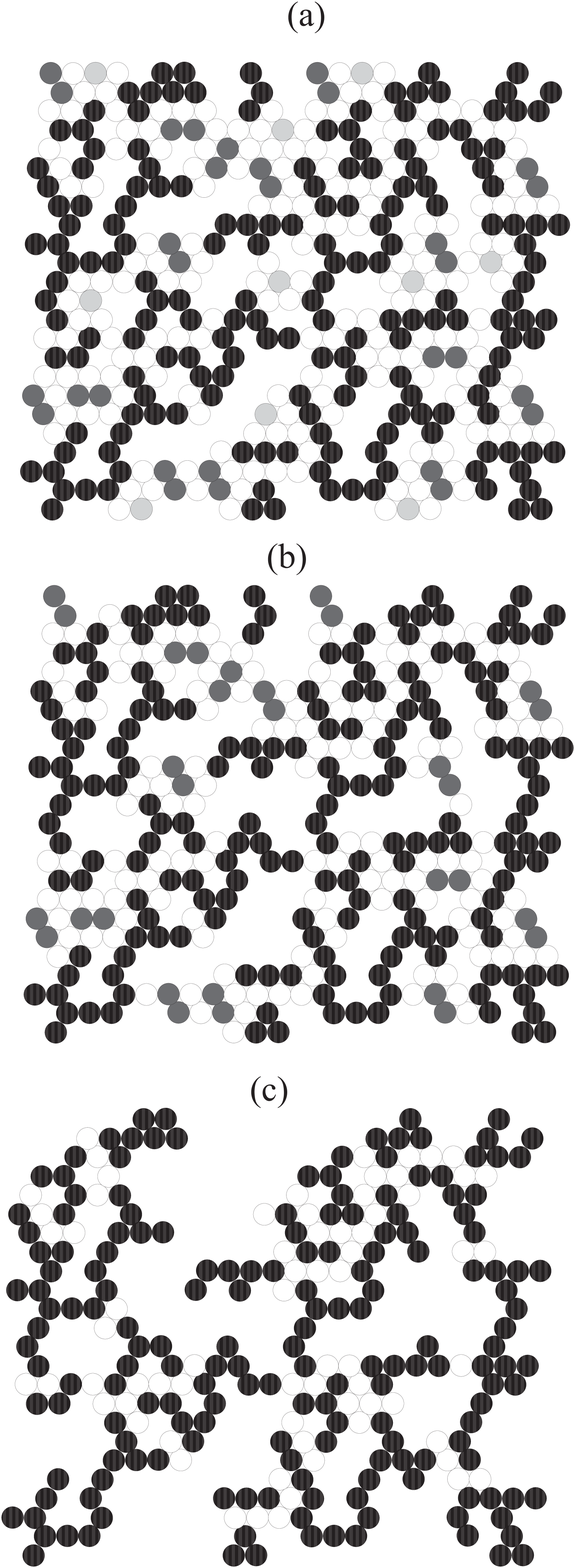}
\caption{An infinite network of NNN-connected clusters. The
smallest clusters (consisting of one grain) are shown light grey;
the second smallest ones (of two grains) are shown dark grey; all
larger metallic clusters are shown black. (a) The full system, (b)
The system with excluded smallest clusters, (c) The system with
excluded both smallest and next smallest clusters. Note that
NNN-percolation exists in all three cases.
 }
 \label{clusters1}
\end{figure}

Let us call two metallic clusters ``NNN-connected'' if there is at
list one link connecting these clusters (or, equivalently, at
least one grain of the first cluster has a next nearest neighbor
belonging to the second cluster). Now fix a certain $\tilde{n}$
and consider the ``refined'' system of all finite clusters with
``large'' sizes $n>\tilde{n}$ (see Fig.\ref{clusters1}b,c). If
$\tilde{n}\gg\xi^{d_f}$ is too large, then only a few of such
clusters will be NNN-connected, and there will be no
``NNN-percolation'' through the system. On the other hand, for
$\tilde{n}\sim 1$ the NNN-percolation should definitely exist. We
define $\tilde{n}_c$ to be the value of $\tilde{n}$ at which the
NNN-percolation is first destroyed in the course of the refinement
process. It is natural to expect
\begin{eqnarray}
\tilde{n}_c\sim n_c\sim \xi^{d_f}. \label{hyp}
\end{eqnarray}
On one hand, it is clear that $\tilde{n}_c$ can not be much larger
than $n_c$, since the distance between so extraordinarily large
clusters is exponentially large and they can not be NNN-connected.
On the other hand, below we demonstrate that at $\tilde{n}_c\sim
n_c$ the NNN-percolation still  exists. Indeed,
 the average intercluster distance $R$ in the
system of clusters with $n$ in the range $(1-\Lambda) n_c<n<n_c$:
\begin{eqnarray}
R\sim [N(n_c)\Lambda n_c]^{1/d}\sim \Lambda^{1/d}\xi, \label{hyp6}
\end{eqnarray}
is of the same order of magnitude as the size of critical clusters
$\xi$. (For derivation of \eqref{hyp6} we used the scaling
relation $\tau-1=d/d_f$ between the critical exponents.) For
$\Lambda>\Lambda_{\rm perc}\sim 1$ there is an infinite network of
interpenetrating clusters (interpenetration of two clusters $i$
and $j$ can be defined, say, as a condition $R_{ij}<\xi$). If
$\Lambda-\Lambda_{\rm perc}$ is not very small, then the fraction
of critical clusters belonging to this interpenetrating network is
of order unity.

To estimate the average number $N_{\rm links}(\xi)$ of links
NNN-connecting a typical pair of interpenetrating critical
clusters, we use the following reasoning (very close to the
arguments of Coniglio \cite{Coniglio} used for determination of
the fractal dimension of ``red bonds''). Let us randomly choose
some of the insulating sites (at concentration $\Delta x\gtrsim
x_c-x$) and substitute conducting ones for them. Then the sample
will be converted to a NN-percolating one with a similar value of
correlation radius $\xi$.  The initial critical clusters will
become NN-connected with probability of order unity and form the
infinite cluster in this new system. On the other hand, typically
two adjacent critical clusters become NN-connected when some
insulating grain that was a critical link between these clusters
is replaced by a conducting grain. The probability of such an
event is $\sim\Delta x\cdot N_{\rm links}(\xi)$. Of course, one
can also imagine a situation when the  NN-connection between these
two clusters is established via a ``chain bridge'' involving one
or more smaller clusters, but the corresponding probabilities are
proportional to higher powers of $\Delta x$ and can be neglected.
So, we conclude that the NN-connection of adjacent critical
clusters is typically established when $\Delta x\cdot N_{\rm
links}(\xi)\sim 1$ and, consequently,
\begin{eqnarray}
N_{\rm links}(\xi)\sim 1/(x_c-x)\gg 1. \label{hyp6}
\end{eqnarray}
 Thus, the
interpenetrating critical clusters are typically multiply
NNN-connected, and therefore the infinite network of
interpenetrating critical clusters is NNN-connected.

Thus, if we exclude all clusters with $n<\tilde{n}(T)$, the
NNN-percolation in the system of remaining clusters persists if
$\tilde{n}(T)<n_c$, or, equivalently, if $T>E_C^{\rm (cr)}$.

Although the conductivity $\sigma$ of a ``refined'' system of
clusters with sizes $n>\tilde{n}$ is nonzero for
$\tilde{n}<\tilde{n}_c$ (since the NNN-percolation is not
destroyed), the magnitude of $\sigma$ is, of course, suppressed by
the refinement. For $1\ll\tilde{n}<\tilde{n}_c$ this suppression
is strong:
\begin{eqnarray}
 \sigma_{\rm ins}(x,\tilde{n})\sim \sigma_{\rm ins}^{(0)}(x)
\tilde{n}^{-\theta}. \label{freezing3a}
\end{eqnarray}
The new critical index $\theta$ can be related to the known ones:
\begin{eqnarray}
 \theta= \frac{(d-2)\nu+s-1}{\nu d_f}.\label{freezing3ak}
\end{eqnarray}
To derive the relation \eqref{freezing3ak} we note that at
$\tilde{n}\sim n_c$ our system consists of critical clusters only,
each pair of adjacent clusters being NNN-connected by $\sim N_{\rm
links}(\xi)$ links, so that we can estimate
\begin{eqnarray}
 \sigma_{\rm ins}(x,\tilde{n}\sim n_c)\sim gN_{\rm links}(\xi)\xi^{2-d}. \label{freezing3appo}
\end{eqnarray}
Then, comparing \eqref{freezing3appo} with \eqref{freezing3a} at
$\tilde{n}\sim n_c$, we immediately arrive at the relation
\eqref{freezing3ak}.  In two dimensions, making use of $\mu=s$ and
Grassberger's precise result~\cite{Gra} $\mu/\nu = 0.9826(8)$, we
obtain $\theta = 0.1227(4)$.  Similarly, in 3D one has
$\theta\approx 0.28$, which would be interesting to verify
numerically.  The index $\theta$ increases with $d$ and reaches
$\theta=1/2$ at critical dimension $d=6$.

\begin{figure}
\includegraphics[width=0.9\columnwidth]{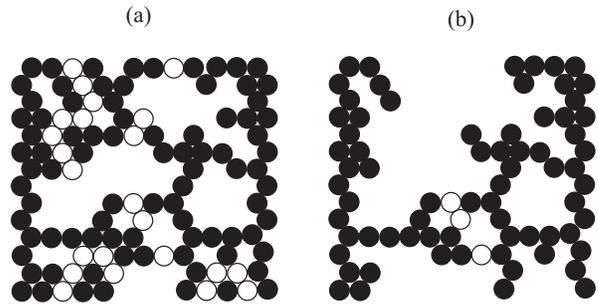}
\caption{A finite critical non-percolating system in the bus-bar
geometry. (a) Before any refinement: all NNN-connected clusters
and corresponding links (white circles) are shown. (b) After
ultimate refinement: only the two bar-clusters and critical links
between them are left.
 }
 \label{bars}
\end{figure}

To verify the assumptions used in the above qualitative
derivation, we performed moderate-statistics numerical
computations for an ensemble of finite critical 2d systems (at
$x=x_c$) in the bus-bar geometry (see Fig.\ref{bars}). The sites
adjacent to the two opposite sides of any sample of the ensemble
belong to the conducting subsystem by definition, while other
sites are conducting with probability $x_c$ and insulating with
probability $1-x_c$. All the conducting clusters with
$n<\tilde{n}$ are removed, but the bar-clusters, connected to the
contacts, are ``protected''---they are never removed.  The current
is allowed to flow from the left bar to the right one via large
clusters with $n>\tilde{n}$, eventually jumping from cluster to
cluster through the links. One half of these systems (those with
the two bars directly connected to each other, so that the current
actually avoids links) is percolating and has very large
conductance $\propto G$. The non-percolating systems of the other
half have conductances $\propto g$ dominated by links. The
expression for the size-dependent conductivity $\sigma_{\rm cr}$,
averaged over the non-percolating half of ensemble,  is
\begin{eqnarray}
 \sigma_{\rm cr}(L,\tilde{n})\sim
gL^{s/\nu}\tilde{n}^{-\theta}, \qquad 1\ll \tilde{n}\ll L^{d_f},
\label{freezing3axx}
\end{eqnarray}
consistent with \eqref{freezing3a}.  The refinement process
typically stops at $\tilde{n}\sim L^{d_f}$, when the size of
maximal removed cluster reaches the systems size $L$ and only the
two protected bar-clusters (NNN-connected by $\sim N_{\rm
link}(L)$ links) are left in the system (see Fig.\ref{bars}b).

\begin{figure}
\includegraphics[angle=270,width=1\columnwidth]{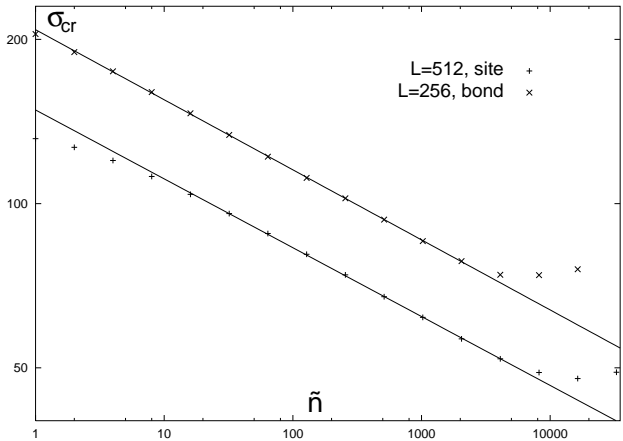}
\caption{Average conductivity of a $L\times L$ square as a
function of the cutoff parameter $\tilde n$ for site and bond
models.  Lines correspond to fits with $\theta=0.128$ (bond) and
$\theta=0.126$ (site).
 }
 \label{x5}
\end{figure}

Averages of finite conductivities $\sigma_{\rm cr}(L,\tilde{n})$
for a $L\times L$ square vs.~cutoff parameter $\tilde n$ are given
in Fig.\ref{x5}. The data shown correspond to $L=256$ for bond
problem and to $L=512$ for site problem. We used the Frank-Lobb
algorithm\cite{FL} to collect statistics over about $2\cdot 10^4$
samples per plot point.  It is obvious that the deviations from
power law \eqref{freezing3axx} are small even for $\tilde n\sim 1$
or $\tilde n$ adjacent to the saturation plateau; standard fits
yield $\theta=0.128(12)$ for the site variant and
$\theta=0.126(10)$ for the bond variant, in good agreement with
each other and with the theoretical prediction.

We also checked numerically that NNN percolation indeed shares the
value $s$ with the ordinary NN problem. Scaling of $\sigma_{\rm
cr}(L,\tilde{n})$ with sample size $L$ for $L=16,32\ldots 1024$
suggested $s/\nu=0.983(9)$ in perfect agreement with percolation
value \cite{Gra} $s/\nu=0.9826(8)$.

The very fact that $\theta\neq 0$, reliably confirmed above both
analytically and numerically, already gives us an important
insight into the physics of the NNN-conduction in an infinite
system with large but finite $\xi$. {\it A priori} one could
imagine two possible scenarios of the conduction process:
\begin{enumerate}
\item The conducting paths  in a non-refined sample consist
predominantly of largest critical clusters with size $n\sim n_c$,
directly connected by critical links. Smaller clusters with $n\ll
n_c$ are irrelevant.  Were this scenario the actual one, the
refinement process would not lead to any considerable suppression
of the conductivity as long as $\tilde{n}\ll n_c$, and,
consequently, one would expect $\theta=0$ and $s=1-(d-2)\nu$.
\item Though the number of relevant conducting paths in a
non-refined sample depends on $x$ and diverges as $x\to x_c$, each
path consists predominantly of small clusters with size $n\sim 1$.
 When all
clusters with $n<\tilde{n}$ are removed from our system due to
refinement, the leading role passes on to the smallest remaining
clusters with $n\sim\tilde{n}$. For this scenario one expects
$\theta>0$ and $s>1-(d-2)\nu$.
\end{enumerate}
Our results definitely indicate that the second scenario, and not
the first one, is real.  Note that the opposition of the two
concepts described above is reminiscent of the alternative between
the oversimplified ``links and nodes'' model \cite{Skal} and a
more sophisticated ``links and blobs'' model \cite{herrmann} in
the standard NN-percolation problem .

\begin{figure}
\includegraphics[width=1\columnwidth]{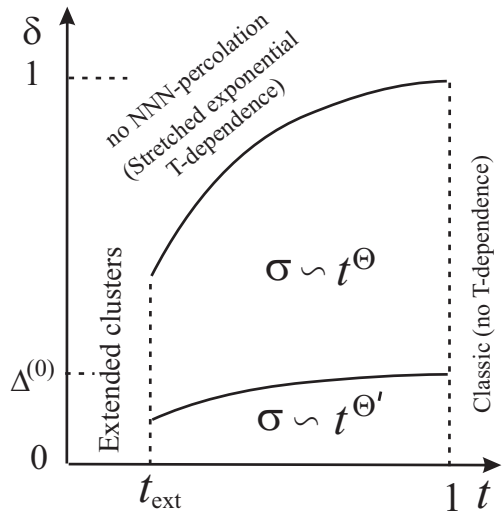}
\caption{Domains of the power-law $T$-dependence on the $\delta
-t$ plane. Here $\delta\equiv x_c-x$ is the distance from the
percolation transition, $t\equiv T/E_C^{(0)}$ is the dimensionless
temperature, $t_{\rm ext}\sim G^{(\nu+s)/[(2-d)\nu+\mu]}\ll 1$.
The upper crossover line is given by $\delta\sim t^{2/(\nu+s)}$,
the lower one---by $\delta\sim \Delta^{(0)}t^{\Theta/(\mu+s)}$.
 }
 \label{phasediagram}
\end{figure}

Thus, substituting \eqref{spread5ptp} into \eqref{freezing3a}, for
the temperature dependence of the conductivity we finally obtain
\begin{eqnarray}
 \sigma_{\rm ins}(x,T)\sim \sigma_{\rm ins}(x,\tilde{n}(T))\sim \sigma_{\rm ins}^{(0)}(x)
 (T/E_C^{(0)})^{\Theta},
\label{freezing3ae}
\end{eqnarray}

\begin{eqnarray}
 \Theta=\frac{d_f\nu\theta}{\nu+s}=\frac{(d-2)\nu+s-1}{\nu +s},
\label{freezing3at}
\end{eqnarray}
valid in the wide intermediate temperature range
\begin{eqnarray}
 E_C^{\rm (cr)}\ll
 T\ll E_C^{(0)},
\label{freezing3at}
\end{eqnarray}
and not very close to the percolation transition (outside the
critical crossover domain):
\begin{eqnarray}
\Delta(T)\ll x_c-x\ll 1,\quad
\Delta(T)=\Delta^{(0)}\cdot(T/E_C^{(0)})^{\Theta/(\mu+s)}.
\nonumber
\end{eqnarray}
Within the critical crossover region (at $|x_c-x|\lesssim
\Delta(T)$) the conductivity stops to depend on $x$:
\begin{eqnarray}
 \sigma_{\rm cross}(x,T)\sim  \sigma_{\rm cross}^{(0)}
(T/E_C^{(0)})^{\Theta'},\nonumber\\
 \Theta'=\Theta \mu/(\mu+s).
\label{freezing3aey}
\end{eqnarray}
The phase diagram for different regimes in the
temperature-concentration plane is shown in
Fig.\ref{phasediagram}.

 At  temperatures $T<E_C^{\rm (cr)}$ the
$T$-dependence of the conductivity becomes stretched exponential;
it is due to the fact that at these low temperatures
$\tilde{n}(T)>n_c$ and the current has no chance to avoid the
Coulomb blockaded clusters. The underlying physics is very
different from the exclusion of the clusters considered above; it
is discussed in a separate publication \cite{feigel}.

In conclusion, we have demonstrated that the Coulomb blockade
effect in a granular material close to the percolation threshold
leads to the power-law temperature dependence of conductivity,
 valid in a wide intermediate
temperature-range. This dependence arises due to the wide
power-law distribution of  the charging energies of conducting
clusters, characteristic for the critical region near the
percolation threshold. The corresponding exponent is universal, it
depends only on the dimensionality of the system and can be
expressed through known critical exponents of the percolation
theory. It is interesting to note that in three dimensions both
exponents $\Theta=0.38$ and $\Theta'=0.28$ are close to the
exponent $1/3$ used for the fitting of the experimentally observed
critical $T$-dependence of conductivity of thick amorphous (but,
presumably, not granulated) films of InO and TiN, in which the
superconductivity was suppressed by strong magnetic field
\cite{gantmakher}. It is possible that for the latter systems the
similar power law $\sigma(T)$-dependence is due to the
self-induced inhomogeneity of disordered systems developing at low
temperatures when the system approaches the MIT (see \cite{FIS}).

The Authors are indebted to M.V.Feigelman and V.F.Gantmakher for
valuable discussions of physics of granular metals. This work was
supported by RFBR grant 06-02-16533.


\begin{thebibliography}{99}
\bibitem{comment}
In general, the conductance $g_{\rm tun}$ depends on the geometry
of the contact (e.g., on the angle between the two bonds formed by
the two ``terminal'' conducting grains and the intermediate
insulating grain). We neglect this dependence, since it does not
affect the universal characteristics of the process.
\end{thebibliography}
\end{document}